# Spectral broadening in self-assembled GaAs quantum dots with narrow size distribution

Francesco Basso Basset,[1,2,a)] Sergio Bietti,[2] Artur Tuktamyshev,[2] Stefano Vichi,[2] Emiliano Bonera,[2] Stefano Sanguinetti[2]

[1]Dipartimento di Fisica, Sapienza Università di Roma, Piazzale A. Moro 5, I-00185, Roma, Italy
[2]L-NESS and Dipartimento di Scienza dei Materiali, Università di Milano-Bicocca, Via Cozzi 53, I-20125, Milano, Italy

The control over the spectral broadening of an ensemble of emitters, mainly attributable to the size and shape dispersion and the homogenous broadening mechanisms, is crucial to several applications of quantum dots. We present a convenient self-assembly approach to deliver strain-free GaAs quantum dots with size distribution below 15%, due to the control of the growth parameters during the preliminary formation of the Ga droplets. This results in an ensemble photoluminescence linewidth of 19 meV at 14 K. The narrow emission band and the absence of a wetting layer promoting dot-dot coupling allow us to deconvolve the contribution of phonon broadening in the ensemble photoluminescence and study it in a wide temperature range.

## I.  INTRODUCTION

Epitaxial quantum dots (QDs) are widely considered for applications in optoelectronic devices, due to unique properties such as the discreteness of their energy levels, the tunability of the wavelength of operation with size, and the high oscillator strength. Several active areas of research are currently focusing on the use of III-V semiconductor nanostructures for devices whose functionality can be observed up to room temperature, including QD lasers,[1–5] QD infrared photodetectors,[6–8] solar cells,[9–11] and quantum molecules for the generation of THz radiation.[12,13]

One of the key parameters to consider when designing the optical properties of the QD ensemble is the spectral broadening. The main contribution usually comes from the inhomogeneous size dispersion. During the years, several approaches have been developed to fabricate ordered and size-controlled nanostructures, but this often comes at the cost of a more cumbersome fabrication process and lower optical quality in terms of spectral diffusion or radiative efficiency.[14] While a certain degree of size dispersion is intrinsic to the self-assembly process, the results on the widely investigated system of InAs/GaAs Stranski-Krastanow QDs illustrate that the size distribution is influenced by the growth parameters employed during the formation of the nanostructures. In optimal condition the emission linewidth at cryogenic temperatures is reduced below 20 meV.[15]

This linewidth is associated with a uniformity threshold beyond which thermally activated mechanisms of homogeneous broadening can become relevant on the ensemble emission when approaching room temperature. The main contribution is usually attributed to the broad band of acoustic-phonon assisted transitions which outweighs the radiative recombination from the zero-phonon line.[16–18]

Temperature-dependent photoluminescence (PL) is a convenient and valuable characterization tool to quantify phonon-related line broadening. However, at non-cryogenic temperature single-dot PL is often unreliable because of the poor signal-to-noise ratio. While collecting light emission from a large ensemble of dots can overcome this problem and help estimate the average properties of the sample, the large distribution of emission energies due to size dispersion and thermally activated dot-dot coupling can hide this information.

a) Author to whom correspondence should be addressed.  Electronic mail:  francesco.bassobasset@uniroma1.it

In this work, we consider GaAs/AlGaAs QDs grown by droplet epitaxy. Using this technique it was possible to choose a heterostructure with limited strain gradient and compositional disorder inside the dot and to remove the wetting layer, which can mediate charge transfer between different dots through thermal escape and retrapping.[19] Droplet epitaxy also offers independent control over the density and the size of the QDs during the formation of the Ga droplets. We identify a growth regime resulting in a narrow size dispersion with respect to the typical values for a self-assembly process. The ensemble PL was studied in a large temperature interval to characterize the phonon broadening of the exciton line and the thermally activated carrier escape.

## II. EXPERIMENTAL

The QDs were fabricated using the droplet epitaxy approach in a conventional III-V molecular beam epitaxy setup equipped for in-situ reflection high energy electron diffraction. The samples were grown starting from a (001)-oriented GaAs wafer. After the GaAs native oxide was desorbed at 580°C, a GaAs buffer layer was grown to achieve an atomically flat surface, followed by an AlGaAs layer with 30% Al content. This material acts as the potential barrier for the QDs and as the substrate for the deposition of Ga droplets. Ga adatoms were deposited on the AlGaAs surface at a temperature of 300°C with a flux of 0.02 ML/s. The first monolayer reacts with the As-rich c(4x4) reconstructed surface establishing a Ga-stabilized (4x6) reconstruction. The droplets are formed by the remaining Ga coverage. At this step, in order to produce some samples dedicated to a morphological characterization, we fabricated some larger droplets by depositing 2.75 MLs and we removed the sample from the chamber. Instead, the standard process to fabricate QDs completely embedded in an AlGaAs matrix consisted in the deposition of 0.06 MLs to act as a seed for QDs, and then the Ga droplets were exposed to an As beam equivalent pressure of $5 \cdot 10^{-5}$ torr at 150°C for 3 minutes to crystallize into GaAs. Subsequently, the nanocrystals underwent a flash procedure, consisting in 10 minutes at 380°C in an As pressure of $4 \cdot 10^{-6}$ torr. Finally, the QDs were covered with another layer of AlGaAs with 30% Al content, namely 10 nm deposited at low temperature followed by 140 nm at 580°C, and capped with 10 nm of GaAs. After growth, the sample underwent a rapid thermal annealing in nitrogen atmosphere at 750°C for 4 minutes. The last step improves the radiative efficiency and causes modest interdiffusion, yet sufficient to remove the wetting layer.[20]

The surface density and the size distribution of the Ga droplets were probed on the sample without capping using atomic force microscopy (AFM). The AFM was employed and operated in non-contact mode to prevent any unwanted perturbation of the liquid droplets. The images were acquired over regions of 5 x 5 µm$^2$ area with a lateral resolution of 2 nm.

The spectroscopic characterization was performed by means of ensemble PL. The sample was excited above the barrier bandgap by focusing the 532 nm line from a Nd:YAG continuous wave (CW) laser on a spot with a diameter of approximately 80 µm. Assuming a typical density of emitters in the $10^8$–$10^9$ cm$^{-2}$ range, several thousands of QDs were simultaneously excited. The PL signal was dispersed by a 150 g/mm diffraction grating in a 500 mm focal length spectrometer and finally detected by a Peltier-cooled CCD.

The role of the excitation power was investigated in the range from 5 µW to 5 mW. The emission was studied in a large temperature range as well, from 14 to 270 K, using a closed-cycle helium cryostat. The high-energy tail of the radiative recombination from the GaAs buffer and substrate layers was modeled as a decaying exponential function[21] and subtracted so to isolate the contribution due to the emission from the QDs. Using this procedure, the measurement of the integrated PL intensity was quantitatively reproducible within 15%.

## III. RESULTS AND DISCUSSION

As described in the previous section, a specific set of growth conditions was chosen to promote the formation of the Ga droplets. Previous studies[22,23] have shown that nanometric Ga droplets self-assemble in a broad space of deposition parameters. Substrate temperatures between 200 and 450°C in combination with Ga fluxes between 0.01 and 1 ML/s create suitable seeds for the subsequent crystallization of GaAs QDs. These parameters directly control the areal density and volume of both the droplets and the QDs, since the position and size of the nanostructures are preserved during As incorporation and crystallization.[24,25] In this work, we consider the combination of an intermediate substrate temperature (300°C) and a low Ga flux (0.02 ML/s). An AFM scan of a sample of Ga droplets grown under these conditions is shown in Fig. 1(a). The surface density of droplets is $7 \cdot 10^8$ cm$^{-2}$. Already from this image it is possible to appreciate that, despite the random ordering typical of self-assembled growth schemes, the size dispersion is low. The normalized, to the average, droplet radius distribution is reported in Fig. 1(b). The distribution is peaked with 15% full width at half maximum (FWHM).

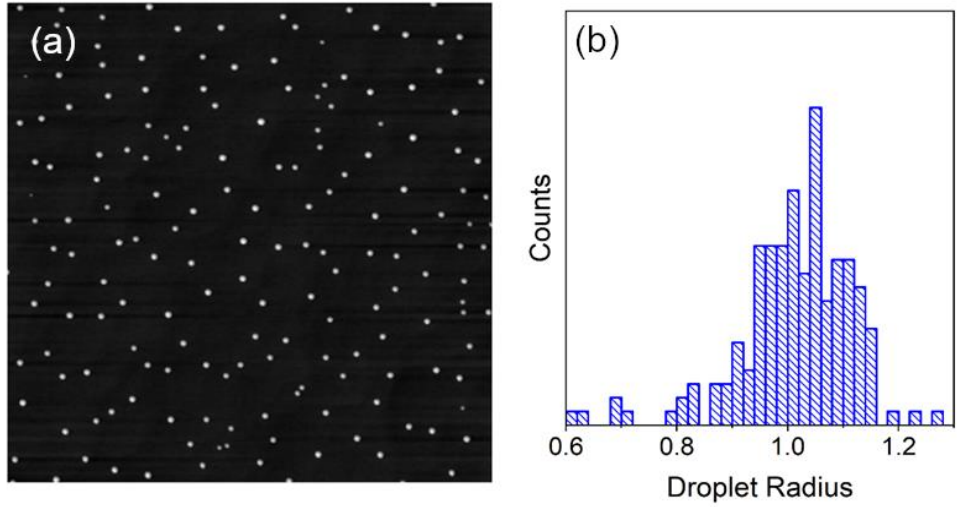

**FIG. 1.** (a) AFM image performed on a 5 x 5 µm² area of the Ga droplets deposited at 300°C with a Ga flux of 0.02 ML/s. (b) Histogram reporting the statistical distribution of the normalized droplet radius.

The size uniformity is expected to be retained after the crystallization into GaAs QDs and the capping with the AlGaAs layer. In fact, well defined atomic interfaces with limited intermixing have been observed for GaAs/AlGaAs QDs grown with droplet epitaxy on a (001)-oriented surface.[26,27] The control over the geometry of the QDs is directly related to the optical properties of their ensemble.

Figure 2(a) shows the PL spectrum collected from the capped sample at 14 K applying 50 µW of laser excitation power, in the spectral window where the emission is attributed to the QDs. In such conditions of low temperature of the sample and low excitation power, the emission peak can be solely attributed to the energy distribution of the s-shell exciton lines, its spectral broadening is inhomogeneous and due to the different confinement potential of different QDs. As expected from the size uniformity unveiled by the AFM measurements on the Ga droplets, the emission peak is spectrally narrow and symmetric. The FWHM is only 19 meV, a value that equals the state-of-the-art results obtained for self-assembled Stranski–Krastanow InAs QDs.[15] In fact, an estimation of the QD size distribution FWHM can be extracted from the PL measurements. The QD electronic confinement energy has a dependence on the QD radius of the form $E_Q = (E_{PL} - E_g) \propto S^{-\alpha}$, with α ranging from 2 (particle in a box) to 1[28], $E_g$ the energy gap of the quantum dot material, and $E_{PL}$ the average emission energy of the QD ensemble. Therefore, the relative radius dispersion of the QD ensemble, *dS*, can be derived from the relative emission energy FWHM, *dE*, by the relation

$$\frac{dS}{S} \approx \frac{1}{\alpha}\frac{dE}{E_Q}$$

Being $E_g$ = 1.515 eV, $E_{PL}$ = 1.66 eV and *dE* = 19 meV, *dS/S* from spectroscopic data ranges from 7% to 13%. These values, while consistent with the AFM measurements, are anyway lower than the actual droplet size dispersion, possibly due to the major role played by Ga diffusion during droplet crystallization in the determination of the QD radius.[29]

Such a low value of inhomogeneous broadening allows to spectrally separate the features of the ensemble emission related to the ground and the first excited exciton state. The radiative recombination from the p-shell states of the QDs distinctly appears by increasing the excitation power, as shown in Fig. 2(b).

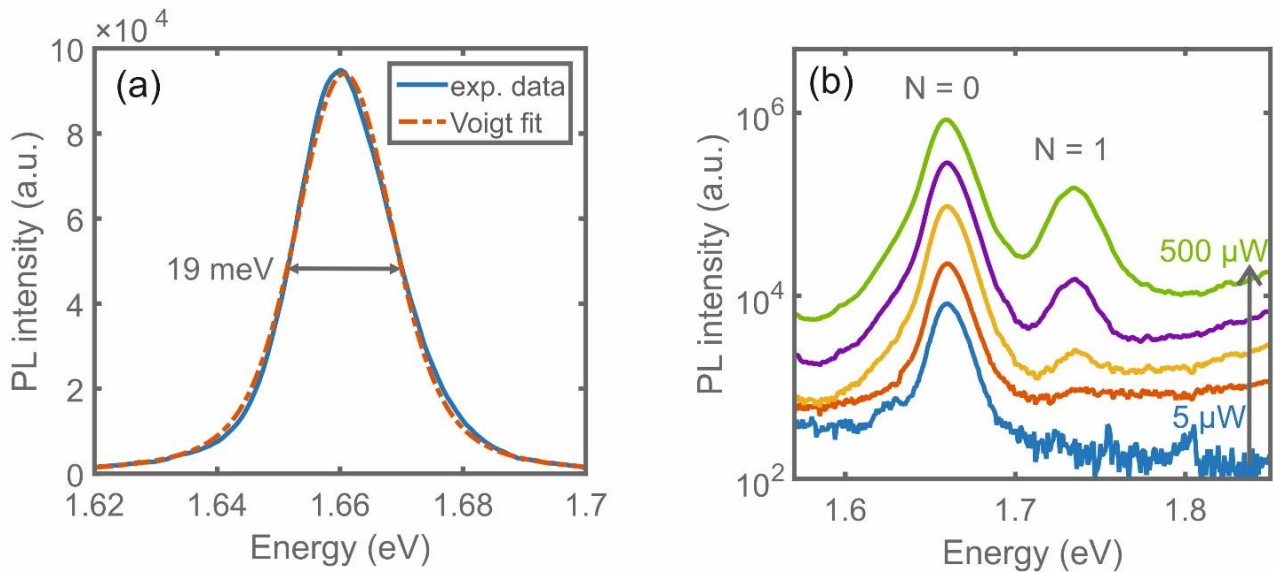

**FIG. 2.** (a) PL spectrum of the emission from an ensemble of QDs, measured at 14K with low excitation power. The FWHM of the emission energy distribution, obtained from a Voigt fit, is reported in the graph. (b) Dependence of the PL spectrum on the laser power. The recombination from the ground and the first excited exciton states are labeled as N = 0 and 1 respectively.

Due to the narrow size dispersion and the absence of a wetting layer, temperature dependent measurements offer insight on single QD properties which are usually hidden by the inhomogeneous broadening in large area optical measurements. The emission from the QDs was observed under a low excitation power of 50 µW in a large temperature range from 14 K to 270 K, close to room temperature. Some of the spectra acquired at different temperatures are reported in Fig. 3(a). As the temperature is increased, the transition energy monotonously redshifts in agreement with Varshni's law.[30] The experimental data is compared in Fig. 3(b) with the Varshni's law assuming the same thermal coefficients as derived for GaAs[31] and an energy offset due to the confinement energy. This behavior is expected for the electronic structure of a single QD and can be observed in ensemble PL when thermally activated dot–dot coupling is negligible. In presence of a wetting layer, instead, the occupation number of a QD with a specific confinement energy is influenced by charge transfer, resulting in a sigmoidal dependence of the peak energy of the ensemble PL on the temperature of the sample.[19,32]

Figure 3(c) shows the temperature dependence of the integrated PL signal from the s-shell states of the QDs. Even if a strong quenching due to the thermally activated carrier escape is present, the emission from the QDs is observed almost up to room temperature. The slope of the curve in the high temperature limit is related to the binding energy of the carriers confined in the QD, that is the activation energy required for the escape and the thermalization outside of the dot. While in this case the asymptotic behavior is not unambiguously identified, an approximate lower bound of (290 ± 40) meV for the activation energy is estimated from the two measurements which are closer to room temperature. This value is near to the binding energy of the exciton, which corresponds to the energy difference of 320 meV between the bandgap recombination peak of the AlGaAs barrier and the centroid of the QD emission in the PL spectrum. Conversely, the independent binding energy of the less strongly bound charge carrier, the heavy hole, is only approximately 150 meV, according to a simple effective-mass single-band model.[33,34] The presence of strong electron-hole correlations up to room temperature is also suggested by the linear dependence of the QD integrated PL intensity on the laser power density,[35] measured up to 210 K as reported in Fig. 3(d).

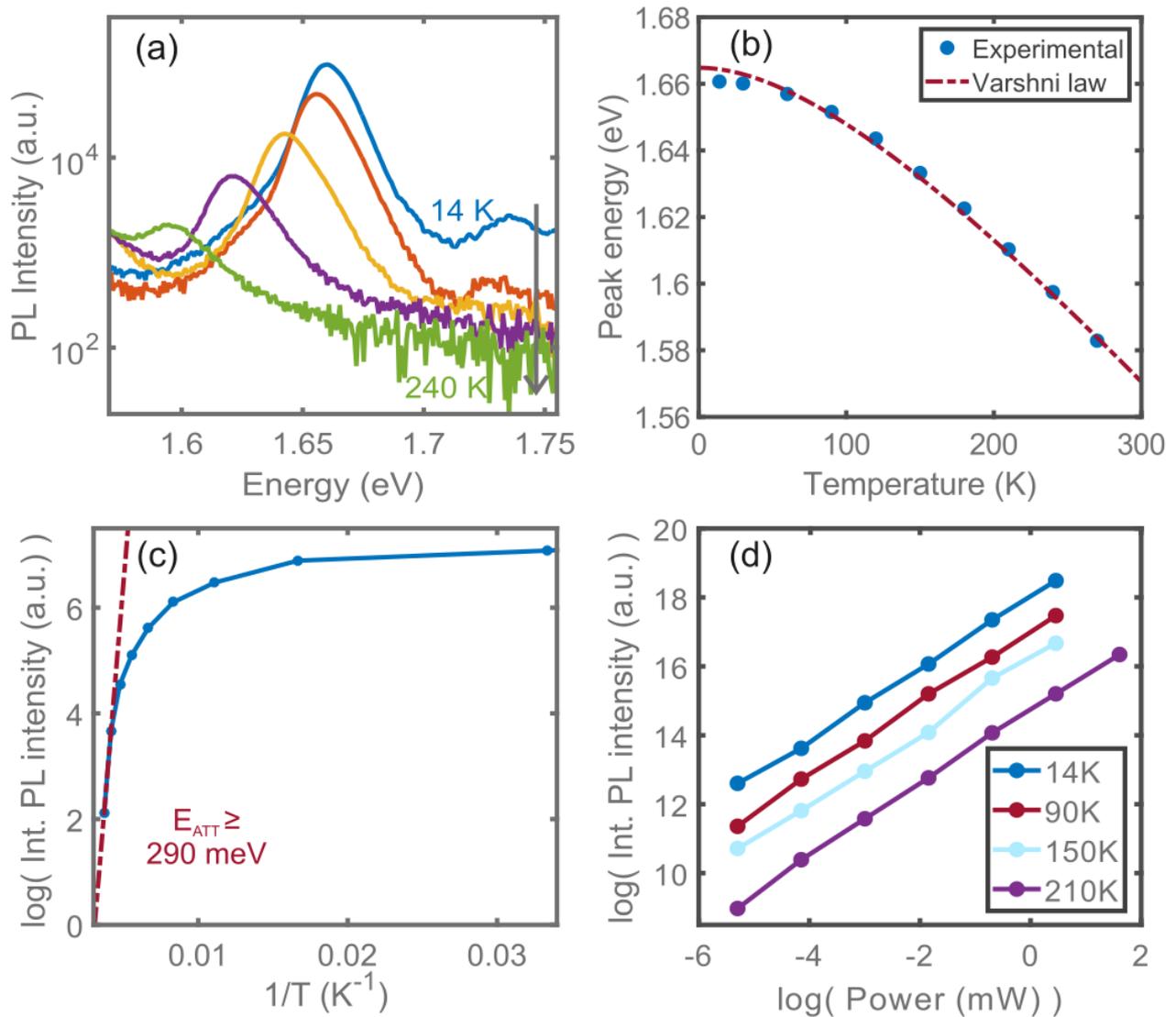

**FIG. 3.** (a) Ensemble PL spectra of the QDs emission at low excitation power at different temperatures of the sample. (b) Peak position of the QDs emission as a function of the temperature of the sample (blue dots) fitted according to the Varshni's law with GaAs thermal coefficients (red dashed line). (c) Arrhenius plot of the integrated PL intensity of the QD ground state transition. The lower bound slope of the signal quenching at high temperature is plotted with the corresponding activation energy. (d) Dependence of the integrated PL intensity from the QDs on the laser excitation power, measured at different temperatures of the sample.

Most notably, the narrow emission peak and the absence of dot-dot coupling allow to study the average single dot line broadening as the temperature increases. The inhomogeneous contribution coming from the size dispersion does not depend on temperature, hence the increase in linewidth shown in Fig. 4(a) is due to a thermally activated mechanism. To quantify this effect, the QD PL spectra were fitted with a Voigt function to extract the FWHM of their Gaussian and Lorentzian components.[36] The Gaussian contribution does not significantly vary with temperature and can be attributed to the inhomogeneous energy distribution of the emitters. Instead, the Lorentzian part broadens as the temperature rises.

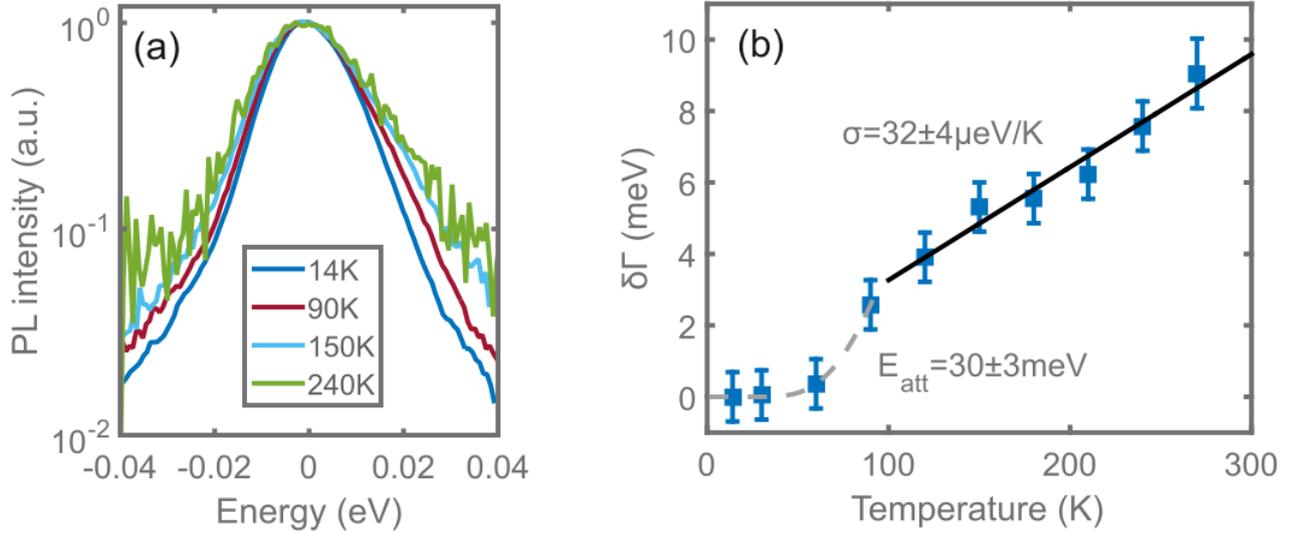

**FIG. 4.** a) Ensemble PL spectra acquired at different temperatures normalized in intensity and centered in energy, so to underscore the difference in spectral broadening. b) Broadening of the PL emission band as a function of the temperature of the sample. Along with the experimental data (blue squares), an exponential fit at low temperatures (dashed grey line), interpreted as the zero-phonon line broadening, and a linear fit at high temperatures (black line), attributed to the contribution of the phonon sidebands.

The increase of the FWHM of the QD emission band with respect to the value measured at 14 K is reported in Fig. 4(b) and can be directly attributed to the phonon-exciton interaction. In the temperature range below 100 K, the energy resolution obtained by the deconvolution of the ensemble emission band is insufficient to get accurate information about the linewidth broadening. This regime has been widely investigated by single dot spectroscopy, and it is described by the broadening of the zero-phonon line that follows a thermally activated behavior,[37] usually attributed to exciton-photon interaction via virtual excitations to higher confined states.[38,39]

More relevant to our scope is the region above 100 K, which is hardly accessible by single dot spectroscopy due to the decrease in brightness and to the spectral spread, often requiring more complex techniques such as four-wave mixing on stacked quantum dots.[40] A handful of experimental studies assessing phonon broadening at non-cryogenic temperatures in different QD systems are listed in Table I. In this regime, the quantum dot emission on the broad phonon sidebands, caused by the coupling of the exciton state with the continuum of acoustic phonons,[17] dominates over the zero-phonon line.[16,41] The data obtained from our analysis show a linear increase of the width of the phonon sidebands characterized by a coefficient of (32 ± 4) µeV/K. This behavior is consistent with previous works on other materials systems[40] where this dependence has been interpreted as a result of a cut-off of the acoustic-phonon modes associated with the inverse of the localization length of the exciton.

If the linear dependence traced in Fig. 4(b) is extrapolated to 300 K a phonon broadening of 9.5 meV is obtained, a relevant figure of merit for the various applications of GaAs/AlGaAs three-dimensionally confined systems at room temperature.

TABLE I. Comparison of the single QD emission line broadening in the high temperature regime dominated by the phonon sidebands for different QD systems reported in the literature.

| Materials system[a] | Spectroscopy method | Reference | σ (μeV/K)[b] | $\Gamma_{RT}$ (meV)[c] |
|---|---|---|---|---|
| DE GaAs/Al$_{0.3}$Ga$_{0.7}$As | Ensemble PL | This work | 32 ± 4 | 9.5 |
| SK In$_{0.7}$Ga$_{0.3}$As/GaAs | 4-wave mixing | Borri (2001)[40] | 20 | 6.4 |
| SK In$_{0.6}$Ga$_{0.4}$As/GaAs | Single dot PL | Bayer (2002)[37] | 20–33 | 3.2–5.3 |
| SK In$_{0.5}$Ga$_{0.5}$As/GaAs | Single dot PL | Matsuda (2001)[42] | 39 | 12 |
| DE InGaAs/GaAs | Ensemble PL | Gurioli (2005)[36] | 29 | 5.7 |
| DEtch GaAs/Al$_{0.45}$Ga$_{0.55}$As | Single dot PL | Benyoucef (2006)[43] | 46 | 8.5 |
| NW-QD GaN/Al$_{0.8}$Ga$_{0.2}$N | Single dot PL | Holmes (2016)[44] | 190 | 19 |
| MEE CdSe/ZnSSe/MgS | Single dot PL | Arians (2007)[45] | 95 | 25.2 |
| CQDs CdSe/PMMA | 3-pulse photon echo | Goupalov (2001)[46,47] | 170 | 62 |

[a]Growth technique used to fabricate the QDs and materials forming the heterostructure (well/barrier). SK stands for Stranski-Krastanow, DE for droplet epitaxy, DEtch for droplet etching, NW-QD for quantum dot in a nanowire, MEE for migration enhanced epitaxy, CQDs for colloidal quantum dots.
[b]Linear coefficient which describes the temperature dependence of the single QD emission line broadening in the high temperature regime dominated by the phonon sidebands. If not explicitly given in the original reference, it is estimated from the linear fit of the linewidth broadening data above 100 K.
[c]Linewidth of the single QD emission at 300 K. If not given in the original reference, it is extrapolated from the linear fit of the linewidth broadening data above 100 K.

## IV. CONCLUSIONS

We have addressed the control over size dispersion in a process of QDs self-assembly, namely droplet epitaxy in the strain-free GaAs/AlGaAs (001) system. We identified a growth regime, characterized by a low Ga flux (0.02 ML/s) and an intermediate substrate temperature (300°C), in which the Ga droplets present a narrow radius distribution with 15% FWHM. The choice over the conditions of droplet formation does not place any strong restriction on the design of the QDs, since their average size and their shape are almost independently set by the total quantity of Ga deposited and by the growth parameters during As incorporation respectively.

The narrow size distribution results in an ensemble PL linewidth at low temperature (14 K) of 19 meV, a value sufficiently low to spectrally resolve the radiative recombination from the s-shell and the p-shell of the QDs. The PL emission from the QDs was investigated in a wide temperature range, from 14 to 270 K, and gave useful information for the application of GaAs/AlGaAs QDs in devices operating close to room temperature. The redshift with increasing temperature follows the Varshni's law, ruling out the presence of thermally activated dot-dot coupling. The quenching rate of the PL signal due to the thermal escape and its linear dependence on the excitation power suggest the presence of electron-hole correlation up to high temperature. The contribution due to homogeneous broadening is singled out from the ensemble PL of the QDs. It is attributed to the phonon sidebands of the QDs ground state transition and quantitatively described by a linear temperature dependence.

### ACKNOWLEDGEMENTS

This project has received funding from the European Union's Horizon 2020 research and innovation programme under the Marie Skłodowska-Curie project 4PHOTON (grant agreement No 721394).